\documentclass[preprint,prd,10pt,aps,showpacs,showkeys,amssymb,nofootinbib]{revtex4-1}

\usepackage{amsmath, amsthm}
\usepackage{color}

\newcommand{\nl}{\nonumber\\ }
\newcommand{\pd}{\partial}
\newcommand{\Tint}[1]{\int\limits_{#1}^\infty\mathrm d}
\newcommand{\Li}{{\rm Li}}
\newcommand{\HF}[3]{F\left(#1,#2;#3;r^2\right)}
\newcommand{\HFgen}[4]{F\left(#1,#2;#3;#4\right)}

\begin{document}

\title{Simple way to the high-temperature expansion of relativistic Fermi-Dirac integrals}

\author{A.S. Khvorostukhin}
\email{hvorost@theor.jinr.ru}
\affiliation{Joint Institute for Nuclear Research,
 141980 Dubna, Russia}
\affiliation{ Institute of Applied Physics, Moldova Academy of Science,
MD-2028 Kishineu, Moldova}

\begin{abstract}
The pressure of an ideal relativistic Fermi gas is computed as an infinite series for high temperatures at nonzero chemical potentials. The expansion of the particle number density, scalar density, and entropy density as first derivatives of the pressure is also found.
\end{abstract}

\maketitle

\section{Introduction}
The properties of relativistic quantum ideal gases are a basic tool for studying more complicated equations of states (EoS), which very often include a sum of ideal gases. Usually, various thermodynamic quantities are expressed in an integral form or as a series of modified Bessel functions (see below). Unfortunately, the integrals cannot be evaluated exactly while the series is fast converged only in the low-temperature limit. So one has to employ numerical integration which is quite a slow procedure. An attempt to hasten numerical calculations was made in \cite{JEL}, but the proposed method is hardly scaled and uses the nonlinear equation solving, which is not simple either. Also, numerical schemes are not allowed to make qualitative conclusions while some authors need high-temperature expansion in a foreseeable form to explore, for example, the chiral transition in QCD \cite{Ohnishi}.

Many authors have tried to obtain the high-temperature expansion (see \cite{HaberWeldon0, Klajn} and reference therein). The first terms can be easily derived (see for example \cite{KG_2006}). For an ideal Bose gas, the task was completely solved by Haber and Weldon \cite{HaberWeldon0}. Also, for the sum of a particle and its antiparticle, the high-temperature expansion of the thermodynamic potential, which is just the pressure up to a factor, was recently obtained in terms of special functions of a complex argument by Klajn \cite{Klajn}.
However, sometimes one needs to separate particles and antiparticles and so it is necessary to know odd power terms in the series. Also the final result \cite{Klajn} for fermions calls for calculations of complex special functions which is not easy. So it is convenient to exclude evidently complex numbers and to give the high-temperature expansion in a similar form as in \cite{HaberWeldon0}.
Comparing \cite{HaberWeldon0} and \cite{Klajn} we see that the method used in the first paper is more complicated. Here we will provide how to get the same result for an ideal Bose gas and the expansion for Fermi gases in a simple way like in \cite{Klajn} without complex numbers in the final expressions.

The paper is organized as follows. In Sec. II, we introduce
the integral to be studied, note how it gives the low-temperature expansion, and obtain the hight-temperature expansion in terms of polylogarithms, $\Li_s(z)$. Our technique is to use the expansion of modified Bessel functions and to perform resummation of the obtained double series. In Sec.~III, the final result is formulated. In Sec. IV, we give the expansion for first derivatives of the pressure. We have included some relevant mathematical information in Appendixes A and B. Appendix C contains the proof of equivalence for my and Klajn's  \cite{Klajn} results for fermions.

\section{Low- and high-temperature expansion in terms of special functions}
We consider the problem of calculation of the pressure of an ideal Bose or Fermi gas. All other quantities can be obtained from the pressure (see Sec. IV).

We start with the integral representation
\begin{eqnarray}
\label{Pdef}
P(T,\mu,m)
&=&-\alpha\, T\int\frac{{\rm d}^3k}{(2\pi)^3} \ln\left[1-\alpha\,e^{\beta(\mu-E)}\right]\,,
\end{eqnarray}
where  $E=\sqrt{k^2+m^2}$ and $m$ is the mass of particles,
$\mu$ is the chemical potential, $T$ is the temperature, $\beta = 1/T$, and the statistics are
\begin{equation}
\alpha=\left\{
\begin{array}{ll}
-1\quad \text{for fermions},\\
+1\quad \text{for bosons}.
\end{array}
\right.
\end{equation}
Integration of Eq. (\ref{Pdef}) by parts results in
\begin{eqnarray}
P(T,\mu,m)
&=&\frac{1}{6\pi^2}\Tint{0}k\, \frac{k^4}{E}\,\frac1{e^{\beta(E-\nu)}-\alpha}\nl
&=&\frac{1}{6\pi^2}\Tint{m}E\,
k^{3}\,\frac1{e^{\beta(E-\nu)}-\alpha}\,.
\end{eqnarray}

It is convenient to introduce dimensionless variables:
\begin{eqnarray}
\lambda=\frac{m}T,\quad \nu=\frac{\mu}T,\quad
r=\frac{\mu}m=\frac{\nu}\lambda\,.
\end{eqnarray}
Then Eq. (\ref{Pdef}) can be rewritten as
\begin{eqnarray}
I_P(\lambda,\nu)&=&\frac{P(T,\mu,m) }{T^4}\\
\label{P_Eint}
&=&\frac{1}{6\pi^2}\, \Tint{\lambda}x\,(x^2-\lambda^2)^{3/2} \,
\frac1{e^{x-\nu}-\alpha}\,.
\end{eqnarray}

Expanding the occupation number for $\nu<\lambda$ so that $e^{\nu-x}<1$,
\begin{eqnarray}
\frac1{e^{x-\nu}-\alpha}=\sum_{k=1}^\infty \alpha^{k+1}e^{k(\nu-x)},
\end{eqnarray}
Equation (\ref{P_Eint}) leads to the well-known result,
\begin{eqnarray}
\label{KseriesF}
I_P(\lambda,\nu)=\frac{\lambda^2}{2\pi^2}\sum\limits_{k=1}^\infty\alpha^{k+1}\frac{K_2(k\lambda)}{k^2}\,e^{k\nu},
\end{eqnarray}
where $K_n(x)$ is the modified Bessel function.
This expansion is valid for any $\lambda\geq0$ and $\nu<\lambda$, but it is quickly converging only for $\lambda\gg 1$.
So it is useful to have a series which can be applied at small $\lambda$.

For this reason we substitute the series representation of the modified Bessel function~\cite{Gradshteyn},
\begin{eqnarray}
K_2(z)&=&\frac2{z^2}-\frac12\nl
&&+\sum_{n=0}^\infty\frac{1}{n!\,(n+2)!}\left(\frac
z2\right)^{2n+2}\Big[\frac{\psi(n+1)+\psi(n+3)}2-\ln\frac
z2\Big],
\end{eqnarray}
and using the polylogarithm function definition (\ref{plogdef}) with Eq. (\ref{dLidnu}) from Appendix B, one easily obtains the following common formula\footnote{Compare with Eq. (22) from \cite{Klajn}.},
\begin{eqnarray}
\label{LiseriesF} I_P(\lambda,\nu)
&=&\frac{\alpha}{\pi^2}\,\Bigg\{\Li_4(\alpha e^\nu)-\frac{\lambda^2}4\,\Li_2(\alpha e^\nu)\nl
&&-\frac{\lambda^2}2\ln\frac{\lambda}2\sum_{n=0}^\infty\frac{1}{n!\,(n+2)!}\left(\frac\lambda2\right)^{2n+2}\Li_{-2n}(\alpha e^\nu)+\nl
&&+\frac{\lambda^2}2\sum_{n=0}^\infty\frac{1}{n!\,(n+2)!}\left(\frac{\lambda}2\right)^{2n+2}\left[\frac{\psi(n+1)+\psi(n+3)}{2}\,\Li_{-2n}(\alpha e^\nu)\right.\nl
&&\left.+\left.\frac{\pd}{\pd s}\Li_s(\alpha e^\nu)\right|_{s\,=\,-2n}\right]\Bigg\}
\end{eqnarray}
which is proved for $\nu<0$ when $|\alpha e^{\nu}|<1$ and is extended for any sign of $\nu$ by analytic continuation. One should note that the expansion (\ref{LiseriesF}) as a whole is not analytic.

A private case is the expression for $\lambda=0$,
\begin{eqnarray}
I_P(0,\nu)=\frac{\alpha}{\pi^2}\,\Li_4(\alpha e^\nu).
\end{eqnarray}

We also note that the nonrelativistic (low-temperature) limit \cite{manuals} is just a consequence of Eq. (\ref{KseriesF}). If one uses the asymptotic series,
\begin{eqnarray}
  K_2(z) &\simeq&  e^{-z}\sqrt{\frac{\pi}{2z}}\sum_{n=0}^\infty\frac{\Gamma(5/2+n)}{\Gamma(5/2-n)n!}\frac{1}{(2z)^n}\,,
\end{eqnarray}
the result is
\begin{eqnarray}
I_P(\lambda,\nu)&=&\alpha\left(\frac{\lambda}{2\pi}\right)^{3/2}\sum_{n=0}^\infty\frac{\Gamma(5/2+n)}{\Gamma(5/2-n)n!}\frac{\Li_{n+5/2}\left(\alpha e^{\tilde\nu}\right)}{(2\lambda)^n}\\
&=&\alpha\left(\frac{\lambda}{2\pi}\right)^{3/2}\Li_{\,5/2}\left(\alpha e^{\tilde\nu}\right)+\ldots,\nonumber
\end{eqnarray}
where $\tilde\nu=\nu-\lambda$ is the nonrelativistic chemical potential. Using Eqs. (\ref{plogdef}) and (\ref{liasympt}), one can obtain the corresponding high- ($e^{\tilde\nu}\ll1$) and low-temperature ($e^{\tilde\nu}\gg1$) expansion\footnote{For fermions, $\tilde \nu\in(-\infty,\infty)$ and $e^{\tilde \nu}\in[0,\infty)$. For bosons, since we consider the termodynamical limit, $V\rightarrow\infty$, and do not consider the Bose condensation, $\tilde\nu\in(-\infty,0]$ or $e^{\tilde\nu}\in[0,1]$. So there is no low-temperature expansion for bosons.}.

\section{The high-temperature expansion through elementary functions.}
Equation (\ref{LiseriesF}) immediately gives us the leading terms of the  high-temperature expansion up to $m^2/T^2$. For higher power terms we need to deal with $\frac{\pd}{\pd s}\Li_s(\alpha e^\nu)$, which is not clear. So it is convenient to replace all special functions in Eq. (\ref{LiseriesF}) by the corresponding series.

Following \cite{HaberWeldon0}, we break up $I_P$ into pieces that are even and odd in $\nu$:
\begin{eqnarray}
I_P^e(\lambda,\nu)&=& \frac12\Big[I_P(\lambda,\nu)+I_P(\lambda,-\nu)\Big],\nl
I_P^o(\lambda,\nu)&=& \frac12\Big[I_P(\lambda,\nu)-I_P(\lambda,-\nu)\Big],\\
I_P(\lambda,\nu)&=&I_P^e(\lambda,\nu)+I_P^o(\lambda,\nu).\nonumber
\end{eqnarray}

\subsection{High-temperature fermions}
Then, for $\alpha=-1$, substituting Eqs. (\ref{li2m_mez}), (\ref{dLids_mez}), and (\ref{lin_mez}) from Appendix B and changing the summation order, we get
\begin{eqnarray}
\label{IPE_fermi}
I_P^e(\lambda,\nu)&=&\,\frac{7\pi^2}{720}+\frac{\nu^2}{24}+\frac{\nu^4}{48\pi^2}-\frac{\lambda^2}{16}\left(\frac{1}{3}+\frac{\nu^2}{\pi^2}\right)
-\frac{\lambda^4}{32\pi^2}\left(\ln\frac{\lambda}\pi+\gamma_E-\frac34\right)\nl
&&+\frac{\lambda^2}{2}\sum_{k=1}^\infty\frac{(-1)^{k+1}\beta(2k+1)}{\Gamma(k+1)\Gamma(k+3)}\left(\frac{\lambda}{2\pi}\right)^{2k+2}\HF{-k}{-k-2}{\frac12}\,,
\end{eqnarray}
\begin{eqnarray}
\label{IPO_fermi}
I_P^o(\lambda,\nu)
&=&\,\frac{\nu}{\pi^2}\Bigg[\frac{3\zeta(3)}{4}+\frac{\ln2}{6}\,\nu^2-\frac{\ln2}{4}\,\lambda^2\nl
&&+\lambda^2\ln\frac{\lambda}2\sum_{k=1}^{\infty}\frac{(-1)^{k}\beta(2k)}{\Gamma(k)\Gamma(k+2)}\left(\frac\lambda{2\pi}\right)^{2k}\,\HF{1-k}{-k-1}{\frac32}\nl
&&+\frac{\lambda^2}2\sum_{k=1}^\infty(-1)^{k}\beta(2k)\left(\frac{\lambda}{2\pi}\right)^{2k}\nl
&&\ \times\Bigg\{\frac{(2r)^{2k}}{\Gamma(2k+2)}-\frac{(2r)^{2k+2}}{\Gamma(2k+4)}-\sum_{i=0}^{k-1}\frac{(2r)^{2i}}{\Gamma(2i+2)}\frac{\psi(k-i)+\psi(k-i+2)}{\Gamma(k-i)\Gamma(k-i+2)}\nl
&&\quad+\frac{2}{\Gamma(k)\Gamma(k+2)}\left[\frac{\beta\,'(2k)}{\beta(2k)}-\ln\pi\right]\HF{1-k}{-k-1}{\frac32}\Bigg\}\Bigg],
\end{eqnarray}
where we have introduced for brevity
\begin{eqnarray}
\beta(x)&=&\Gamma(x)\zeta(x)(1-2^{-x})
,\\\frac{\beta'(x)}{\beta(x)}&=&\psi(x)+\frac{\zeta'(x)}{\zeta(x)}+\frac{\ln2}{1-2^{-x}}-\ln2\,,
\end{eqnarray}
and the hypergeometric functions $\HF{a}{b}{c}$ are just polynomials:
\begin{eqnarray}
\label{HF_IPE}
\HF{-k}{-k-2}{\frac12} &=&\sum_{i=0}^k\frac{k!\,(k+2)!}{(k-i)!\,(k+2-i)!}\frac{(2r)^{2i}}{(2i)!}\,,\\
\HF{1-k}{-k-1}{\frac32} &=&\sum_{i=0}^{k-1}\frac{(k-1)!\,(k+1)!}{(k-i-1)!\,(k+1-i)!}\frac{(2r)^{2i}}{(2i+1)!}\,.
\end{eqnarray}
$\HF{a}{b}{c}$ are polynomials since $a$ and $b$ are negative integers or zero.

It can be shown that Eq. (\ref{IPE_fermi}) completely coincides with the result \cite{Klajn} for fermions. The corresponding proof is given in Appendix C.

Rewriting $\HF{-k}{-k-2}{\frac12}$ as Jacobi polynomials and applying the recurrence relation, we find that the series representation of $I_P^e(\lambda,\nu)$ is converged for $\lambda+|\nu|<\pi$.


\subsection{High-temperature bosons}
In the same way, for $\alpha=1$, using Eqs. (\ref{li2m_ez}), (\ref{dLids_ez}), and (\ref{lin_ez}) from Appendix B we get for bosons
\begin{eqnarray}
I_P^e(\lambda,\nu)&=&\,\frac{\pi^2}{90}+\frac{\nu^2}{12}-\frac{\nu^4}{48\pi^2}-\frac{\lambda^2}{8}\left(\frac{1}{3}-\frac{\nu^2}{2\pi^2}\right)+\frac{(\lambda^2-\nu^2)^{3/2}}{12\pi}\nl
&&+\frac{\lambda^4}{32\pi^2}\left(\ln\frac{\lambda}{4\pi}+\gamma_E-\frac34\right)\nl
&&+\lambda^2\sum_{k=1}^\infty\frac{(-1)^kb(2k+1)}{\Gamma(k+1)\Gamma(k+3)}\left(\frac\lambda{4\pi}\right)^{2k+2}\HF{-k}{-k-2}{\frac12}\,,
\end{eqnarray}
\begin{eqnarray}
I_P^o(\lambda,\nu)&=&\,\frac{\nu}{\pi^2}\Bigg\{\zeta(3)-\frac{7}{24}\,\lambda^2+\frac{11}{36}\,\nu^2+\frac{(\lambda^2-\nu^2)^{3/2}}{6}\,\frac{\arcsin r}{\nu}\nl
&& +\ln\frac\lambda2\left[\frac{\lambda^2}{4}-\frac{\nu^{2}}{6}\right.\nl
&&\quad\left.
-\lambda^2\sum_{k=1}^\infty\frac{(-1)^kb\,(2k)}{\Gamma(k)\Gamma(k+2)}\left(\frac\lambda{4\pi}\right)^{2k}\,\HF{1-k}{-k-1}{\frac32}\right]\nl
&&
-\frac{\lambda^2}2\sum_{k=1}^\infty(-1)^kb(2k)\left(\frac\lambda{4\pi}\right)^{2k}\nl
&&\times\Bigg\{\frac{(2r)^{2k}}{\Gamma(2k+2)}-\frac{(2r)^{2k+2}}{\Gamma(2k+4)}\,-\sum_{i=0}^{k-1}\frac{(2r)^{2i}}{\Gamma(2i+2)}\frac{\psi(k-i)+\psi(k+2-i)}{\Gamma(k-i)\Gamma(k+2-i)}\nl
&&+\frac{2}{\Gamma(k)\Gamma(k+2)}\left[\frac{b\,'(2k)}{b\,(2k)}-\ln(2\pi)\right]\HF{1-k}{-k-1}{\frac32}\Bigg\}\Bigg\},
\end{eqnarray}
which exactly coincide with the result \cite{HaberWeldon0}. Here,
\begin{eqnarray}
b(x)&=\Gamma(x)\zeta(x).
\end{eqnarray}
This exercise allows us to be sure of the correctness of the result for the Fermi-Dirac integral. 

\section{Other thermodynamical quantities}
Besides the pressure, three other quantities are often used: the particle number density, the scalar density, and the entropy density which are the first derivatives of the pressure,
\begin{eqnarray}
n(T,\mu,m)&=&\frac{\pd P(T,\mu,m)}{\pd\mu}=T^3\frac{\pd I_P(\lambda,\nu)}{\pd\nu}=T^3I_n(\lambda,\nu),\\
\rho^{sc}(T,\mu,m)&=&-\frac{\pd P(T,\mu,m)}{\pd m}=-T^3\frac{\pd I_P(\lambda,\nu)}{\pd\lambda}=T^3I_{sc}(\lambda,\nu),\\
s(T,\mu,m)&=&\frac{\pd P(T,\mu,m)}{\pd T}=T^3\left[4I_P(\lambda,\nu)+\lambda\,I_{sc}(\lambda,\nu)-\nu\, I_n(\lambda,\nu)\right]\\
&=&T^3I_s(\lambda,\nu)\nonumber.
\end{eqnarray}
As a result, from the first law of thermodynamics, one obtains the energy density
\begin{eqnarray}
\varepsilon(T,\mu,m)&=&Ts(T,\mu,m)+\mu\, n(T,\mu,m)-P(T,\mu,m)\nl
&=&T^4\left[3I_P(\lambda,\nu)+\lambda\,I_{sc}(\lambda,\nu)\right].
\end{eqnarray}

Below we list the complete high-temperature expansion of the Fermi-Dirac integrals for the particle number density,
\begin{eqnarray}
I_n^e(\lambda,\nu)
&=&\frac{3\,\zeta(3)}{4\pi^2}+\frac{\nu^2\ln2}{2\pi^2}-\frac{\lambda^2\ln2}{4\pi^2}\nl
&&+4\ln\frac{\lambda}2\sum_{k=1}^{\infty}\frac{(-1)^{k}\beta(2k)}{\Gamma(k)\Gamma(k+2)}\left(\frac\lambda{2\pi}\right)^{2k+2}\,\HF{1-k}{-k-1}{\frac12}\nl
&&+2\sum_{k=1}^\infty(-1)^{k}\beta(2k)\left(\frac{\lambda}{2\pi}\right)^{2k+2}G^n_k(r)\,,
\end{eqnarray}
\begin{eqnarray}
G^n_k(r)
&=&\frac{(2r)^{2k}}{\Gamma(2k+1)}-\frac{(2r)^{2k+2}}{\Gamma(2k+3)}-\sum_{i=0}^{k-1}\frac{(2r)^{2i}}{\Gamma(2i+1)}\frac{\psi(k-i)+\psi(k-i+2)}{\Gamma(k-i)\Gamma(k-i+2)}\nl
&&+\frac{2}{\Gamma(k)\Gamma(k+2)}\left[\frac{\beta\,'(2k)}{\beta(2k)}-\ln\pi\right]\,\HF{1-k}{-k-1}{\frac12}\,,
\end{eqnarray}
\begin{eqnarray}
I_n^o(\lambda,\nu)
&=&\nu\Bigg[\frac{1}{12}+\frac{\nu^2}{12\pi^2}-\frac{\lambda^2}{8\pi^2}\nl
&&+2\sum_{k=1}^\infty\frac{(-1)^{k+1}\beta(2k+1)}{\Gamma(k)\Gamma(k+2)}\left(\frac{\lambda}{2\pi}\right)^{2k+2}\HF{1-k}{-k-1}{\frac32}\Bigg],
\end{eqnarray}
the scalar density,
\begin{eqnarray}
I_{sc}^e(\lambda,\nu)
&=&\,\frac{1}{24}+\frac{\nu^2}{8\pi^2}+\frac{\lambda^2}{8\pi^2}\left(\ln\frac{\lambda}\pi+\gamma_E-\frac12\right)\nl
&&+\sum_{k=1}^\infty(-1)^{k}\frac{\beta(2k+1)}{\Gamma(k+1)\Gamma(k+2)}\left(\frac{\lambda}{2\pi}\right)^{2k+2}\HF{-k}{-k-1}{\frac12}\,,
\end{eqnarray}
\begin{eqnarray}
I_{sc}^o(\lambda,\nu)& =&\frac{\nu}{\pi^2}\Bigg[\frac{\ln2}{2}\nl
&&-2\ln\frac{\lambda}2\sum_{k=1}^{\infty}(-1)^{k}\frac{\beta(2k)}{\Gamma(k)\Gamma(k+1)}\left(\frac\lambda{2\pi}\right)^{2k}\,\HF{1-k}{-k}{\frac32}\nl
&&-\sum_{k=1}^\infty(-1)^{k}\beta(2k)\left(\frac{\lambda}{2\pi}\right)^{2k}G^{sc}_k(r)\Bigg]\,,
\end{eqnarray}
\begin{eqnarray}
G^{sc}_k(r)& =&\frac{(2r)^{2k}}{\Gamma(2k+2)}-\sum_{i=0}^{k-1}\frac{(2r)^{2i}}{\Gamma(2i+2)}\frac{\psi(k-i)+\psi(k-i+1)}{\Gamma(k-i)\Gamma(k-i+1)}\nl
&&+\frac{2}{\Gamma(k)\Gamma(k+1)}\left[\frac{\beta'(2k)}{\beta(2k)}-\ln\pi\right]\,\HF{1-k}{-k}{\frac32},
\end{eqnarray}
and the entropy density,
\begin{eqnarray}
I_s^e(\lambda,\nu)&=&\,\frac{7\pi^2}{180}+\frac{\nu^2}{12}-\frac{\lambda^2}{24}+\frac{\lambda^4}{32\pi^2}\nl
&&+\lambda^2\sum_{k=1}^\infty\frac{(-1)^{k}\beta(2k+1)}{\Gamma(k)\Gamma(k+3)}\left(\frac{\lambda}{2\pi}\right)^{2k+2}\HF{-k}{-k-2}{\frac12},
\end{eqnarray}
\begin{eqnarray}
I_s^o(\lambda,\nu)&=&\,\frac{\nu}{\pi^2}\Bigg[\frac{9\zeta(3)}{4}-\frac{\lambda^2}{4}\left(1-\frac{2r^2}{3}\right)\ln2\nl
&&+\lambda^2\ln\frac{\lambda}2\sum_{k=1}^{\infty}(-1)^{k+1}\frac{(2k-1)\beta(2k)}{\Gamma(k)\Gamma(k+2)}\left(\frac\lambda{2\pi}\right)^{2k}\,\HF{1-k}{-k-1}{\frac32}\nl
&&+\frac{\lambda^2}2\sum_{k=1}^\infty(-1)^{k}(2k-1)\beta(2k)\left(\frac{\lambda}{2\pi}\right)^{2k}G^s_k(r)\Bigg]\,,
\end{eqnarray}
\begin{eqnarray}
G^s_k(r)&=&\frac{(2r)^{2k+2}}{\Gamma(2k+4)}-\frac{(2r)^{2k}}{\Gamma(2k+2)}+\sum_{i=0}^{k-1}\frac{(2r)^{2i}}{\Gamma(2i+2)}\frac{\psi(k-i)+\psi(k-i+2)}{\Gamma(k-i)\Gamma(k-i+2)}\nl
&&-\frac{2}{\Gamma(k)\Gamma(k+2)}\left[\frac{\beta'(2k)}{\beta(2k)}+\frac1{2k-1}-\ln\pi\right]\HF{1-k}{-k-1}{\frac32}.
\end{eqnarray}
All functions $G^a_k(r)$ are polynomials.

\section{Summary}
The high-temperature ($\lambda\rightarrow0$) expansion of the integral (\ref{Pdef}) for Fermi-Dirac statistics has been obtained. For thermodynamic applications, this result allows us to obtain all thermodynamic quantities. The proposed method for obtaining high-temperature expansion is simple and reproduces the results of \cite{HaberWeldon0,Klajn}.

\section*{Acknowledgements}
We are very grateful to A. Parvan and V.D. Toneev for discussions and valuable remarks. I am also very grateful to the referee for many important improvements and the idea of Appendix C.

\appendix
\section{Properties of gamma, digamma, and Riemann zeta functions}
In our consideration, we often meet three special functions: gamma, digamma, and Riemann zeta functions. In this Appendix we just quote some useful relations for these functions.

First of all, we have \cite{Gradshteyn}
\begin{align}
\label{gammanext}
\Gamma(x+1)&=x\,\Gamma(x),\\
\label{gammaneg}
\Gamma(x)\,\Gamma(1-x)&=\frac{\pi}{\sin\pi x}\,,\\
\label{gammadouble}
\Gamma(2x)&=2^{2x-1}\frac{\Gamma(x)\,\Gamma(x+1/2)}{\Gamma(1/2)}\,.
\end{align}
Then applying Eq. (\ref{gammaneg}) to the nominator and denominator, we obtain
\begin{align}
\label{gammarel}
\frac{\Gamma(-x-n)}{\Gamma(-x)}&=(-1)^{n}\frac{\Gamma(x+1)}{\Gamma(x+n+1)}.
\end{align}
Also we permanently keep in mind that \cite{Gradshteyn}
\begin{align}
\Gamma(n)&=(n-1)!\,,\quad \Gamma(1/2)=\sqrt{\pi}\,,
\end{align}
where $n\in\mathbb{N}$.

In our computations we need to know how to take the derivative of the gamma function. So we deal with
\begin{align}
\label{dgamma}
\Gamma'(x)&=\psi(x)\Gamma(x)
\end{align}
and we use the following features of the digamma function \cite{Gradshteyn},
\begin{align}
\label{psineg}
\psi(1-z)&=\psi(z)+\pi\cot \pi z\,,\\
\label{psione}
\psi(1)=-\gamma_E,
\end{align}
where $\gamma_E$ is Euler's constant. From Eqs. (\ref{psineg}) and (\ref{gammaneg}) it can be found that
\begin{align}
\label{psigammarel}
\frac{\psi(-n)}{\Gamma(-n)}&=(-1)^{n+1}\Gamma(n+1)\,,\quad n=0,1,2,\ldots
\end{align}

The last remarkable object which occurs in the paper is the Riemann zeta function, $\zeta(x)$. We use that~\cite{Gradshteyn}
\begin{align}
\label{zetaneg}
\zeta(z)&=2(2\pi)^{z-1}\,\Gamma(1-z)\,\zeta(1-z)\sin\frac{\pi z}2
\end{align}
and
\begin{align}
\label{zetafornegint}
\zeta(0)=-\frac12\,,\quad\zeta(-2n)&=0,\quad\zeta(1-2n)=(-1)^n\frac{2\Gamma(2n)\,\zeta(2n)}{(2\pi)^{2n}}\,,\nl
\zeta'(0)&=-\frac12\,\ln(2\pi)\,,
\end{align}
where $n\in\mathbb{N}$.

Combining Eqs. (\ref{gammarel}) and (\ref{psigammarel})--(\ref{zetafornegint}) and l'H\^{o}pital's rule, one can find the limit~\cite{Robinson},
\begin{align}
&\lim_{s\rightarrow n}\left[\Gamma(1-s)(-z)^{s-1}+\frac{\zeta(s-n+1)}{\Gamma(n)}\,z^{n-1}\right]\nl
&=\frac{z^{n-1}}{\Gamma(n)}\lim_{x\rightarrow 0}\frac{\frac{\Gamma(x+1)}{\Gamma(x+n)}\,\Gamma(n)(-z)^{x}+2(2\pi)^{x}\,\zeta(-x)\sin\frac{\pi (x+1)}2}{\frac1{\Gamma(-x)}}\nl
\label{limit}
&=\frac{z^{n-1}}{\Gamma(n)}\left[\gamma_E+\psi(n)-\ln(-z)\right].
\end{align}
Also differentiating both sides of Eq. (\ref{zetaneg}), we get
\begin{align}
\label{dzetaneg}
\zeta'(-2n)&=(-1)^{n}\frac{\Gamma(2n+1)\,\zeta(2n+1)}{2(2\pi)^{2n}}\,,\\
\label{dzetaneg1}
\zeta'(1-2n)&=(-1)^{n+1}\frac{2\Gamma(2n)\,\zeta(2n)}{(2\pi)^{2n}}\Bigg[\psi(2n)+\frac{\zeta'(2n)}{\zeta(2n)}-\ln(2\pi)\Bigg]
\end{align}
for $n\in\mathbb{N}$.

For brevity, we sometimes also use the Dirichlet eta function
\begin{align}
\eta(z)&=\left(1-2^{1-z}\right)\,\zeta(z)
\end{align}
whose properties follow from  the properties of $\zeta(x)$.


\section{Properties of polylogarithms}
The polylogarithm function can be defined by power series \cite{HaberWeldon0},
\begin{align}
\label{plogdef}
\Li_s(z)&=\sum_{k=1}^\infty\frac{z^{k}}{k^s},
\end{align}
where $z,s\in\mathbb{C}$ and $|z|<1$. It can be extended to $|z| \geq 1$ by the procedure of analytic continuation.

From the definition, one immediately obtains
\begin{align}
\label{Lideriv}
\frac{\pd\, \Li_s(z)}{\pd z}&=\frac{\Li_{s-1}(z)}z
\end{align}
and
\begin{align}
\label{dLidnu}
\frac{\pd\, \Li_s(z)}{\pd s}&=-\sum_{k=1}^\infty\frac{z^{k}}{k^s}\,\ln k.
\end{align}

We need a series expansion of $\Li_s(e^{z})$ and $\Li_s(-e^{z})$ for small $z$. The first one is obtained in \cite{Robinson}, and we quote the result:
\begin{align}
\label{lis_ez}
\Li_s(e^{z})&=\Gamma(1-s)(-z)^{s-1}+\sum_{k=0}^\infty\frac{\zeta(s-k)}{k!}\,z^k,\quad |z|\leq2\pi,\ s\not\in \mathbb{N}\,.
\end{align}
For an integer $s=n$, one has to take the limit of Eq.(\ref{lis_ez})\footnote{See Eq. (\ref{limit}).} and the result is \cite{HaberWeldon0,Robinson}
\begin{align}
\label{lin_ez}
 \Li_{n}(e^{z})&=\frac{z^{n-1}}{\Gamma(n)}\Big[\psi(n)+\gamma_E-\ln(-z)\Big]+\sum_{k=0,\,k\neq
n-1}^\infty\frac{\zeta(n-k)}{k!}\,z^k\nl
&=\sum_{k=0}^{n-2}\frac{\zeta(n-k)}{k!}\,z^{k}+\frac{z^{n-1}}{(n-1)!}\Big[\psi(n)+\gamma_E-\ln(-z)\Big]-\frac{z^{n}}{2\,n!}\nl
&+2z^{n-1}\sum_{k=1}^\infty(-1)^{k}\frac{\Gamma(2k)\,\zeta(2k)}{\Gamma(2k+n)}\,\left(\frac{z}{2\pi}\right)^{2k},\quad
n\in\mathbb{N},
\end{align}
where Eq. (\ref{zetafornegint}) is used. 

To obtain the expansion for $\Li_s(-e^{z})$, one should just apply Eq. (\ref{lis_ez}) for $z + i\pi$ or use Taylor's theorem and Eq. (\ref{Lideriv}). 
Then it results in \cite{Wood}
\begin{align}
\label{lis_mez}
\Li_s(-e^{z})&=-\sum_{k=0}^\infty\frac{\eta(s-k)}{k!}\,z^k,\quad |z|<\pi,\ s\in \mathbb{C}.
\end{align}
For comparison with Eq. (\ref{lin_ez}), we separately write the expansion for $s=n$:
\begin{align}
\label{lin_mez}
 \Li_{n}(-e^{z})&=-\frac{z^{n-1}}{(n-1)!}\,\ln2-\sum_{k=0,\,k\neq
n-1}^\infty\frac{\eta(n-k)}{k!}\,z^k\nl
&=-\sum_{k=0}^{n-2}\frac{\eta(n-k)}{k!}\,z^k-\frac{z^{n-1}}{(n-1)!}\,\ln2-\frac{z^{n}}{2\,n!}\nl
&+2z^{n-1}\sum_{k=1}^\infty(-1)^{k}\frac{(1-2^{-2k})\Gamma(2k)\,\zeta(2k)}{\Gamma(2k+n)}\left(\frac{z}{\pi}\right)^{2k},\quad
n\in\mathbb{N}.
\end{align}
For convenience, we also separately give the expressions for $s=-2m$, $m=0,1,2,\ldots$,
\begin{align}
\label{li2m_ez}
\Li_{-2m}(e^z)&=-\frac{\Gamma(2m+1)}{z^{2m+1}}-\frac12\,\delta_{m0}\nl&
+\frac{2}{(2\pi)^{2m+1}}\sum_{k=m+1}^\infty(-1)^k\frac{\Gamma(2k)\,\zeta(2k)}{\Gamma(2k-2m)}\,\left(\frac{z}{2\pi}\right)^{2k-2m-1}\,,\\
\label{li2m_mez}
\Li_{-2m}(-e^z)&=-\frac12\,\delta_{m0}\nl&+\frac{2}{\pi^{2m+1}}\sum_{k=m+1}^{\infty}(-1)^{k}\frac{\Gamma(2k)\,\zeta(2k)(1-2^{-2k})}{\Gamma(2k-2m)}\left(\frac{z}{\pi}\right)^{2k-2m-1}\,,
\end{align}
which are directly obtained from Eqs. (\ref{lis_ez}) and (\ref{lis_mez}) where Eq. (\ref{zetafornegint}) is taken into account.

Finally, we should find the derivatives of polylogarithms over the index for $s=-2m$. It can be easily made by taking the corresponding derivative on the rhs of Eq. (\ref{lis_ez}),
\begin{align}
\label{dLids_ez}
\left.\frac{\pd}{\pd
s}\Li_{s}(e^z)\right|_{s=-2m}&=-\frac{\ln(2\pi)}{2}\,\delta_{m0}+\frac{\psi(2m+1)\Gamma(2m+1)}{z^{2m+1}}-\frac{\Gamma(2m+1)}{z^{2m+1}}\,\ln(-z)\nl
&-\frac{2}{(2\pi)^{2m+1}}\sum_{k=m+1}^\infty(-1)^k\frac{\Gamma(2k)\zeta(2k)}{\Gamma(2k-2m)}\,\left(\frac{z}{2\pi}\right)^{2k-2m-1}\nl
&\quad\quad\quad\times\left[\psi(2k)+\frac{\zeta'(2k)}{\zeta(2k)}-\ln(2\pi)\right]\nl
&+\frac{1}{2(2\pi)^{2m}}\sum_{k=m+\delta_{m0}}^\infty(-1)^{k}\frac{\Gamma(2k+1)\,\zeta(2k+1)}{(2k-2m)!}\,\left(\frac{z}{2\pi}\right)^{2k-2m},
\end{align}
and of Eq. (\ref{lis_mez})
\begin{align}
\label{dLids_mez}
\left.\frac{\pd}{\pd
s}\Li_{s}(-e^z)\right|_{s=-2m}&=-\frac12\,\delta_{m0}\,\ln\frac{\pi}{2}\nl
&+\frac{1}{\pi^{2m}}\sum_{k=m+\delta_{m0}}^\infty(-1)^{k}\frac{\Gamma(2k+1)\zeta(2k+1)(1-2^{-2k-1})}{\Gamma(2k-2m+1)}\left(\frac{z}{\pi}\right)^{2k-2m}\nl
&-\frac{2}{\pi^{2m+1}}\sum_{k=m+1}^\infty(-1)^{k}\frac{\Gamma(2k)\zeta(2k)(1-2^{-2k})}{\Gamma(2k-2m)}\left(\frac{z}{\pi}\right)^{2k-2m-1}\nl
&\quad\quad\quad\times
\left[\psi(2k)+\frac{\zeta'(2k)}{\zeta(2k)}-\ln(2\pi)\right]\nl
&-\frac{2\ln2}{\pi^{2m+1}}\sum_{k=m+1}^\infty(-1)^{k}\frac{\Gamma(2k)\zeta(2k)}{\Gamma(2k-2m)}\left(\frac{z}{\pi}\right)^{2k-2m-1}.
\end{align}
To obtain the last two expressions, one should use Eqs. (\ref{dzetaneg}) and (\ref{dzetaneg1}).

Let us also note the asymptotic expansion \cite{Wood}
\begin{align}
\label{liasympt}
\Li_s(-e^z)&=-2\sum_{n=0}^\infty\frac{\eta(2n)}{\Gamma(s+1-2n)}\,z^{s-2n}\,.
\end{align}

\section{EQUIVALENCE OF EQ. (\ref{IPE_fermi}) TO KLAJN'S RESULT \cite{Klajn}.}

Applying Eq. (51) of \cite{Klajn} for fermions so that $\beta\tilde\mu = \nu+i\pi$, we obtain
\begin{align}
\frac{\Omega}{2\,T^4}&=\frac{\pi^2}{3}\,B_4\left(\frac12-i\frac{\nu}{2\pi}\right)+\frac{1}{4}\,B_2\left(\frac12-i\frac{\nu}{2\pi}\right)\lambda^2
-\frac{\lambda^4}{32\pi^2}\,\left(\ln\frac\lambda\pi-\frac34-2\ln2\right)\nl
&+\frac{\lambda^4}{32\pi^2}\sum_{k=0}^\infty\frac{(-1)^k}{k!(k+2)!}\left[\psi^{(2k)}
\left(\frac12-i\frac{\nu}{2\pi}\right)+\psi^{(2k)}\left(\frac12+i\frac{\nu}{2\pi}\right)\right]
\left(\frac{\lambda}{4\pi}\right)^{2k}\,.
\end{align}
We need  to prove that this expression is equal to Eq. (\ref{IPE_fermi}).

The Bernoulli polynomials are $B_4(z) = z^4-2z^3 + z^2-1/30$ and $B_2(z) = z^2-z + 1/6$. This gives
\begin{eqnarray}
B_4\left(\frac12-i\frac{\nu}{2\pi}\right)&=&\frac{\nu^4}{16\pi^4}+\frac{\nu^2}{8\pi^2}+\frac7{240}\,,\\
B_2\left(\frac12-i\frac{\nu}{2\pi}\right)&=&-\frac{\nu^2}{4\pi^2}-\frac1{12}\,.
\end{eqnarray}

The derivatives of the polygamma functions may be expanded in a Taylor series about the
argument $1/2$:
\begin{align}
\psi^{(2k)}\left(\frac12-i\frac{\nu}{2\pi}\right)+\psi^{(2k)}\left(\frac12+i\frac{\nu}{2\pi}\right)&=
2\sum_{j=0}^\infty\frac{(-1)^j}{(2j)!}\,\psi^{(2k+2j)}\left(\frac12\right)\left(\frac{\nu}{2\pi}\right)^{2j}.
\end{align}

Collecting the last three expressions, separating out the term $2k+2j = 0$, and changing the sum order, results in
\begin{align}
\label{Omega1}
\frac{\Omega}{2\,T^4}&=\frac{7\pi^2}{720}+\frac{\nu^2}{24}+\frac{\nu^4}{48\pi^2}
-\frac{\lambda^2}{16}\left(\frac13+\frac{\nu^2}{\pi^2}\right)
-\frac{\lambda^4}{32\pi^2}\,\left(\ln\frac\lambda\pi+\gamma_E-\frac34\right)\nl
&+\lambda^2\sum_{j=1}^\infty(-1)^{j}\psi^{(2j)}\left(\frac12\right)\left(\frac{\lambda}{4\pi}\right)^{2j+2}\sum_{k=0}^j\frac{1}{k!(k+2)!}\frac{1}{(2j-2k)!}\,\left(\frac{\nu}{2\pi}\right)^{2j-2k}
\left(\frac{\lambda}{4\pi}\right)^{2k-2j}\,,
\end{align}
where $\psi\left(\frac12\right)=-\gamma_E-2\ln2$ is used.

Taking into account \cite{Gradshteyn}
\begin{align}
\psi^{(l)}(x)&=(-1)^{l+1}\Gamma(l+1)\sum_{n=0}^\infty\frac1{(x+n)^{l+1}}\,,
\end{align}
one can show
\begin{eqnarray}
\psi^{(l)}\left(\frac12\right)&=-2^{l+1}\beta(l+1)
\end{eqnarray}
for $l\in\mathbb{N}$.

The last formulas together with $\nu=r\lambda$ immediately lead Eq. (\ref{Omega1}) to the form of Eq. (\ref{IPE_fermi}).

\end{document}